\documentclass[aps,prd,twocolumn,showpacs,superscriptaddress]{revtex4-2}
\usepackage{amsmath}   
\usepackage{amssymb}
\usepackage{tabularx}
\usepackage{textcomp}
\usepackage{accents}
\usepackage{isotope}
\usepackage{multirow}
\usepackage{graphicx}  
\usepackage{bm}   
\usepackage{rotating}
\usepackage{tcolorbox}
\usepackage{gensymb}
\usepackage{subcaption}
\usepackage{siunitx}
\usepackage{tabularx}
\usepackage{amstext}
\usepackage{mhchem}
\usepackage{xspace}
\usepackage{booktabs}
\usepackage{hyperref}
\usepackage{inputenc}
\usepackage{placeins}
\usepackage{cleveref}
\usepackage{siunitx}
\usepackage{xcolor}

\newcommand{\costhetasun}{cos\xspace$\theta_{\odot}$} 
\newcommand{\beight}[0]{$^{8}$B }

\newcommand*{\snoplusFlux}{$\left(5.36^{+0.41}_{-0.39}\text{(stat.)}^{+0.17}_{-0.16}\text{(syst.)} \right)\times10^{6}$\,cm$^{-2}$s$^{-1}$}
\newcommand*{\snoplusFluxUnoscillated}{$\left(2.32^{+0.18}_{-0.17}\text{(stat.)}^{+0.07}_{-0.05}\text{(syst.)}\right)\times10^{6}$\,cm$^{-2}$s$^{-1}$}

\newcommand{\LivetimeDSI} {92.1}    
\newcommand{\LivetimeDSII} {190.3}  

\begin{document}
\title{Measurement of the $\ce{^{\textbf{8}}}$B Solar Neutrino Flux Using the Full SNO+ Water Phase Dataset}
\hfill \break
\date{August 27, 2024} 
\author{ A.\,Allega}
\affiliation{\it Queen's University, Department of Physics, Engineering Physics \& Astronomy, Kingston, ON K7L 3N6, Canada}
\author{ M.\,R.\,Anderson}
\affiliation{\it Queen's University, Department of Physics, Engineering Physics \& Astronomy, Kingston, ON K7L 3N6, Canada}
\author{ S.\,Andringa}
\affiliation{\it LIP - Laborat\'{o}rio de Instrumenta\c{c}\~{a}o e  F\'{\i}sica Experimental de Part\'{\i}culas, Av. Prof. Gama Pinto, 2, 1649-003 Lisboa, Portugal}
\author{ M.\,Askins}
\affiliation{\it University of California, Berkeley, Department of Physics, CA 94720, Berkeley, USA}
\affiliation{\it Lawrence Berkeley National Laboratory, 1 Cyclotron Road, Berkeley, CA 94720-8153, USA}
\author{ D.\,M.\,Asner} 
\affiliation{\it Brookhaven National Laboratory, P.O. Box 5000, Upton, NY 11973-500, USA}
\author{ D.\,J.\,Auty}
\affiliation{\it University of Alberta, Department of Physics, 4-181 CCIS,  Edmonton, AB T6G 2E1, Canada}

\author{ A.\,Bacon}
\affiliation{\it University of Pennsylvania, Department of Physics \& Astronomy, 209 South 33rd Street, Philadelphia, PA 19104-6396, USA}
\author{ F.\,Bar\~{a}o}
\affiliation{\it LIP - Laborat\'{o}rio de Instrumenta\c{c}\~{a}o e  F\'{\i}sica Experimental de Part\'{\i}culas, Av. Prof. Gama Pinto, 2, 1649-003 Lisboa, Portugal}
\affiliation{\it Universidade de Lisboa, Instituto Superior T\'{e}cnico (IST), Departamento de F\'{\i}sica, Av. Rovisco Pais, 1049-001 Lisboa, Portugal}
\author{ N.\,Barros}
\affiliation{\it LIP - Laborat\'{o}rio de Instrumenta\c{c}\~{a}o e  F\'{\i}sica Experimental de Part\'{\i}culas, Rua Larga, 3004-516 Coimbra, Portugal}
\affiliation{\it Universidade de Coimbra, 
Departamento de F\'{\i}sica (FCTUC), 3004-516, Coimbra, Portugal}
\author{ R.\,Bayes}
\affiliation{\it Queen's University, Department of Physics, Engineering Physics \& Astronomy, Kingston, ON K7L 3N6, Canada}
% ...

\collaboration{The SNO+ Collaboration}


\begin{abstract}
    The SNO+ detector operated initially as a water Cherenkov detector. The implementation of a sealed covergas system midway through water data taking resulted in a significant reduction in the activity of $^{222}$Rn daughters in the detector and allowed the lowest background to the solar electron scattering signal above 5\,MeV achieved to date. This paper reports an updated SNO+ water phase $^8$B solar neutrino analysis with a total livetime of 282.4 days and an analysis threshold of 3.5\,MeV. The $^8$B solar neutrino flux is found to be \snoplusFluxUnoscillated \,assuming no neutrino oscillations, or \snoplusFlux \,assuming standard neutrino oscillation parameters, in good agreement with both previous measurements and Standard Solar Model Calculations. The electron recoil spectrum is presented above 3.5\,MeV.
\end{abstract}
\maketitle

\section{Introduction}
\label{sec:Introduction}
Neutrinos are produced in the core of the Sun by a variety of nuclear reactions. In the higher energy portion of the solar neutrino spectrum, where water Cherenkov detectors are sensitive, neutrinos from \beight decay dominate the flux \cite{BS2005-OP}. The flux and spectrum of these neutrinos have been measured by a number of experiments \cite{SNO_LETA,SuperKSolar,BX:first_b8,Kamland_solar}. A remaining goal of solar neutrino experiments is to measure the upturn in electron neutrino survival probability that is expected below about 4\,MeV, due to the transition between the vacuum and matter-dominated oscillation regimes. The shape of this transition is sensitive to possible models of new physics \cite{Upturn_Wurm,Upturn_Minakata}.

The SNO+ experiment \cite{SnoplusDetector} operated in its initial phase as a kt-scale water Cherenkov detector. In this phase SNO+ was sensitive to solar neutrino interactions via the neutrino-electron elastic scattering interaction \cite{Bahcall_ES}. The low levels of intrinsic background in SNO+ combined with the large overburden at SNOLAB (6010 m.w.e) have enabled a leading search for ``invisible" modes of nucleon decay \cite{SnoplusNDUpdated}, while upgrades to the detector electronics have allowed the detector to operate with reduced thresholds, enabling SNO+ to efficiently detect neutron captures \cite{SnoplusNeutronProton} and detect reactor antineutrinos \cite{SnoplusAntinu} using pure water. A study of \beight solar neutrinos in SNO+ was previously published \cite{SnoplusPrior8B} using an initial ``commissioning" data set that was collected before the SNO+ sealed covergas system \cite{SnoplusDetector} was brought online, but which nevertheless showed very low background levels. This paper presents a measurement of the \beight solar neutrino flux across the full SNO+ water phase including the previously published dataset and an additional \LivetimeDSII\, live days of data with even lower backgrounds. The very low background level in the post-cover gas dataset allowed an analysis threshold of 3.5 MeV, equal to the lowest so far achieved with the water Cherenkov technique \cite{SNO_LETA,SuperKSolar}. 

\section{Detector}
\label{sec:Detector}
SNO+ is a multipurpose detector that re-purposes much of the hardware from the Sudbury Neutrino Observatory \cite{SNO_NIM}. The target fluid is contained in a 6-metre radius spherical acrylic vessel (AV). While the AV is currently filled with liquid scintillator, it previously housed 0.9 kt of ultrapure water (UPW). Surrounding the vessel is an array of 9362 inward-facing PMTs situated on a stainless steel PMT support structure (PSUP) about 8.3\,m in radius. The detector is suspended within a urylon-lined cavity containing a futher 7 kt of UPW, which provides shielding. The SNO+ detector hardware is further described in \cite{SnoplusDetector}.

For the purposes of this paper, a key aspect of the SNO+ detector is the covergas system, which is described in detail in \cite{SnoplusDetector}. The covergas system was designed to achieve $\lesssim$7\,mBq/m$^3$ of $^{222}$Rn in the head space above the AV. To achieve this, the covergas volume is sealed and static (as opposed to the system in SNO in which boil-off nitrogen was continually flowed through the AV covergas system). To avoid potentially damaging pressure differentials across the AV during air pressure changes in the laboratory, the system incorporates a series of flexible ``bags" and an emergency venting system through a series of ``u-trap" manometers that are partially filled with linear alkyl benzene (LAB). The sealed covergas system was brought online in September 2018, and the level of $^{222}$Rn-supported $^{214}$Bi activity in the SNO+ water was observed to decrease from $10^{-13} - 10^{-14}$\,gU/g \cite{SnoplusNDFirst} to $(5.8\pm 0.7 ^{+1.5}_{-1.3}) \times 10^{-15}$\,gU/g \cite{SnoplusNDUpdated}. 

\section{Data and Data Selection}  
\label{sec:DataSelection} 
The data collected from the SNO+ water phase was categorized into two distinct sets based on the status of the covergas system. The first dataset (``DS-I") was gathered from May through December 2017, with the old covergas system in place, and corresponds to the data previously published. In this analysis \LivetimeDSI\ days of DS-I livetime were used. The subsequent lower background data with the sealed covergas system online is referred to as dataset II (``DS-II"). DS-II was collected between October 2018 and July 2019, and includes \LivetimeDSII\ live-days. 

\subsection{Low Level Cuts and Offline Trigger} \label{sec:LowLevelCuts}
In order to reduce data volume and ensure a uniform, well understood trigger threshold in all data taking phases, an analysis threshold was applied to select events with at least 15 PMT triggers that fall within the 400-ns event window, and with at least 10 of those falling within an 89-ns window. A suite of low-level cuts, identical to those used in \cite{SnoplusPrior8B}, was further applied to reject non-physics events and instrumental backgrounds based on the trigger system and PMT timing information.

\subsection{Event Reconstruction} \label{sec:Reconstruction} 
For events passing the low-level selection criteria, higher-level event characteristics were then evaluated. The position, direction, and timing of an event were estimated using a likelihood fit to the observed photon detection times, under the assumption that all observed light is Cherenkov. Energy was then estimated using the number of triggered PMTs, with corrections of offline PMTs, detector geometry, and optical attenuation applied using the reconstructed position and direction of the event. The same event reconstruction algorithms as in \cite{SnoplusPrior8B}, with the updated optical model, were used in this analysis.

\subsection{High Level Cuts} \label{sec:HighLevelCuts}
Events that were successfully reconstructed were subject to further data selection cuts. Cuts are placed on the isotropy of the PMT signals in each event as quantified by the $\beta_{14}$ parameter \cite{SNO_Salt} and the ``in-time ratio" (ITR), which describes the ratio of number of hit PMTs within a 7.5\,ns ``prompt time window" to all hit PMTs in an event \cite{SnoplusPrior8B}. Additionally, for this analysis several new cuts were introduced based on reconstruction figures of merit (``FOM"), as described in \cite{SnoplusAntinu}, with cut thresholds determined from calibration data. These cuts mainly remove poorly reconstructed events at low energy. A summary of the analysis cuts and their impact on the number of events in the two data periods is given in Table \ref{table:trigger}. The signal sacrifice due to the combined data cleaning cuts is 1.2\%, and is corrected for in the analysis.

\begin{table}
    \setlength{\tabcolsep}{4pt}
\begin{center}
\begin{tabular}{|p{3.1cm}||p{2.2cm}|p{2.2cm}|}
\hline
\hline
    Selection & Events Passing (DS-I) & Events Passing (DS-II)\\
\hline
    Total & 10,083,081,664 & 12,472,093,737 \\
    Low Level & 3,757,559,668 & 7,300,180,917 \\
    Offline Trigger & 122,628,131 & 591,080,758 \\
    Valid Reconstruction & 24,969,085 &  107,685,755 \\
    High Level Cuts  & 6,230,266 & 21,098,131 \\ 
    Energy  & 19,140 & 2,365,169 \\
    Fiducial Volume  & 330 & 932 \\ 
\hline
\hline
\end{tabular}
\caption{Dataset reduction for each applied cut. Note that the energy thresholds and fiducial volume cuts applied were different in the two data sets, as described in the text.}
\label{table:trigger}
\end{center}
\end{table}

\subsection{Fiducial Volume and Analysis Threshold} \label{sec:FVCuts}
In DS-I, the $^{222}$Rn distribution in the detector was observed to be variable and non-uniform as the result of radon ingress down the neck of the AV. To mitigate this, the DS-I data was divided into six distinct periods with similar background levels. For each of these six periods, the fiducial volume was determined based on background rate and distribution. To optimize the fiducial volumes, the background rate as a function of energy and fiducial volume was determined using events in a cos\xspace$\theta_{\odot}$ sideband (\costhetasun \textless 0, which selects events pointing back towards the Sun) in a 10\% subset of the data. The expected number of signal events in a given fiducial volume and energy range was estimated using the expected solar neutrino interaction rate, the fiducial mass, and the livetime of the data sub-period. The expected statistical significance of the extracted solar neutrino signal could then be determined for that fiducial volume, energy range, and data sub-period, allowing the fiducial volume for each energy range and data sub-period to be optimized by maximizing the expected significance of the extracted signal. The resulting optimized fiducial volumes for the different data periods, and their accumulated livetimes, are shown in Table \ref{tb:TBFVCuts}.

The analysis threshold for each data sub-period was determined by selecting the energy range over which the predicted statistical significance of the extracted signal in the optimized fiducial volume was greater than 0.5\,$\sigma$. In this updated analysis, two of the data sub-periods included in the earlier analysis were excluded due to low expected signal significance at all thresholds. 

\renewcommand{\arraystretch}{1.2}
\begin{table}[h!]
\centering
{\begin{tabular}{|c|c|c|c|c|c|}
\hline
Data Set   & $T_{e}$   & R   & z & Live & Exposure\\
    & (MeV)   &(m)  & (m) & Days & (kt-day)\\
\hline \hline
DS-I-a           & \multicolumn{3}{c|}{Dropped from analysis} & 5.0          & - \\ 
\hline
DS-I-b          & \multicolumn{3}{c|}{Dropped from analysis} & 14.8       &  - \\
\hline
DS-I-c          & 5.0 \textless \, $T_{e}$  \textless \, 6.0  & R\textless 4.4 & z \textless 3.5 & \multirow{2}{*}{29.7} & 10.6 \\
           & 6.0 \textless \, $T_{e}$ \textless 15.0 & R\textless 5.3 & -  & & 19.1 \\
\hline
DS-I-d          & 5.0 \textless \, $T_{e}$ \textless \, 6.0 & R\textless 5.0 & -   & \multirow{2}{*}{28.6}   & 15.4\\
           & 6.0 \textless \, $T_{e}$ \textless 15.0 & R\textless 5.3 & -  & & 18.3 \\
\hline
DS-I-e          & 5.0 \textless \, $T_{e}$ \textless \, 6.0  & R\textless 5.3 & z \textless 3.5  & \multirow{2}{*}{11.2}     &   6.6   \\
           & 6.0 \textless \, $T_{e}$ \textless 15.0 & R\textless 5.3 & -  &  &  7.2\\
\hline
DS-I-f          & 5.0 \textless \, $T_{e}$ \textless \, 6.0  & R\textless 5.3 & z \textless 3.5 & \multirow{2}{*}{22.6}  & 13.3 \\   
           & 6.0 \textless \, $T_{e}$ \textless 15.0 & R\textless 5.3 & -  &  & 14.4 \\
      \hline     
DS-II          & 3.5 \textless \, $T_{e}$ \textless \, 5.0  & R\textless 4.4 & - & \multirow{2}{*}{190.3} & 67.7 \\  
           & 5.0 \textless \, $T_{e}$ \textless 15.0 & R\textless 5.3 & -  &  & 118.2 \\

\hline
\end{tabular}}
\caption {Optimized fiducial volumes and exposures for the different data sets. Note that DS-I is sub-divided into six periods based on variations in background rate and distribution, in the same way as in \cite{SnoplusPrior8B}.}
\label{tb:TBFVCuts}
\end{table}


\section{Detector Calibration Using $^{241} \text{Am}^9\text{Be}$}
\label{sec:DetectorCalibration} 

The response of the SNO+ detector is calibrated using a number of deployed calibration sources, as described in detail in previous publications \cite{SnoplusNDFirst, SnoplusPrior8B, SnoplusOptics}. A key feature of the current result is the lower analysis threshold relative to previous SNO+ publications. Data analysis at this lower threshold required validating the detector calibration at lower energies. This was accomplished using data from an $^{241}$Am$^9$Be (AmBe) neutron calibration source. This approach was enabled by the novel ability of SNO+ to efficiently trigger on the 2.2-MeV signal from neutron captures on protons \cite{SnoplusNeutronProton}. Roughly 60\% of neutrons produced by the AmBe source result in an excited state of the $^{12}$C daughter and a subsequent 4.4-MeV de-excitation gamma ray. It was therefore possible to produce a ``tagged" set of calibration events consisting of a prompt 4.4-MeV signal and a delayed 2.2-MeV signal from the subsequent neutron capture - a combination that nicely spans the newly analyzed energy range.

Key detector performance parameters calibrated in this way include the energy scale and resolution and the angular resolution (by using the baseline from the source to the interaction point of the 4.4\,MeV gamma to estimate the recoil electron direction, following the approach in \cite{SNO_8B}). The energy scale and resolution systematics were assessed by simulating each calibration source run and fitting the simulated data, convolved with a gaussian response function, to the corresponding detector data.

\begin{figure}[htbp]
    \centering
\includegraphics[width=0.5\textwidth]{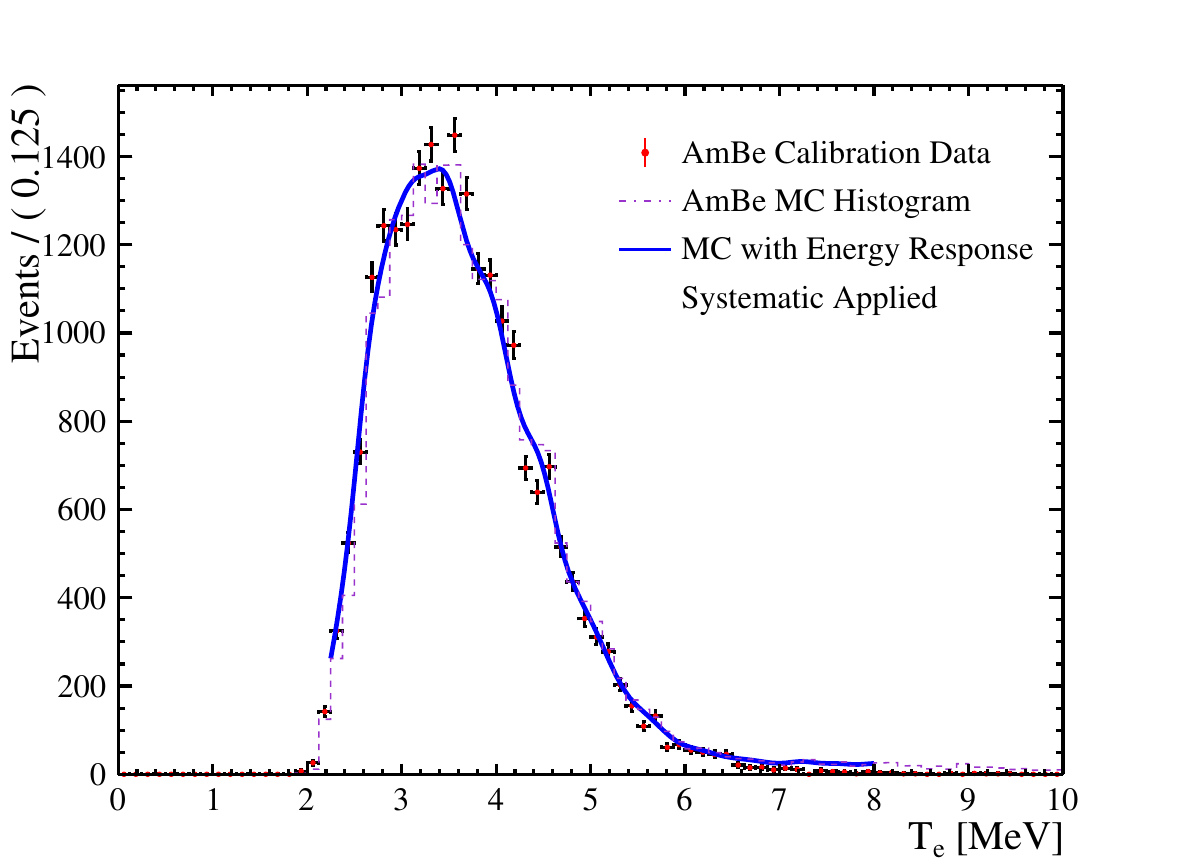}%
\caption{Reconstructed energy of prompt events from the AmBe source at a central run in data and MC simulation. The blue line shows the MC with the best fit Gaussian convolution used to assess the energy response systematics.}
\label{fig:Deployment_AmBe}
\end{figure}

\begin{figure}[htbp]
    \centering
\includegraphics[width=0.48\textwidth]{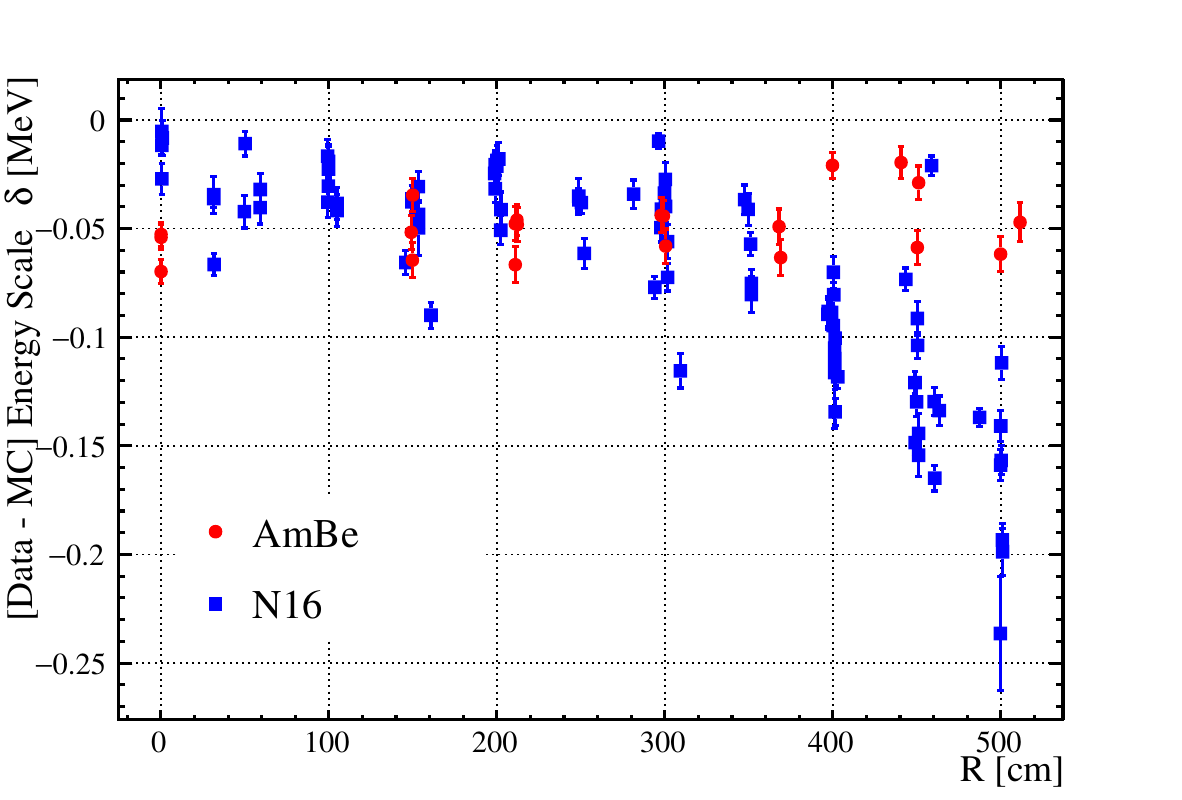}%
\caption{Comparison of the energy scale in Data and MC simulation as a function of calibration source radial position.}
\label{fig:Deployment_EScale}
\end{figure}

\begin{figure}[htbp]
    \centering
\includegraphics[width=0.48\textwidth]{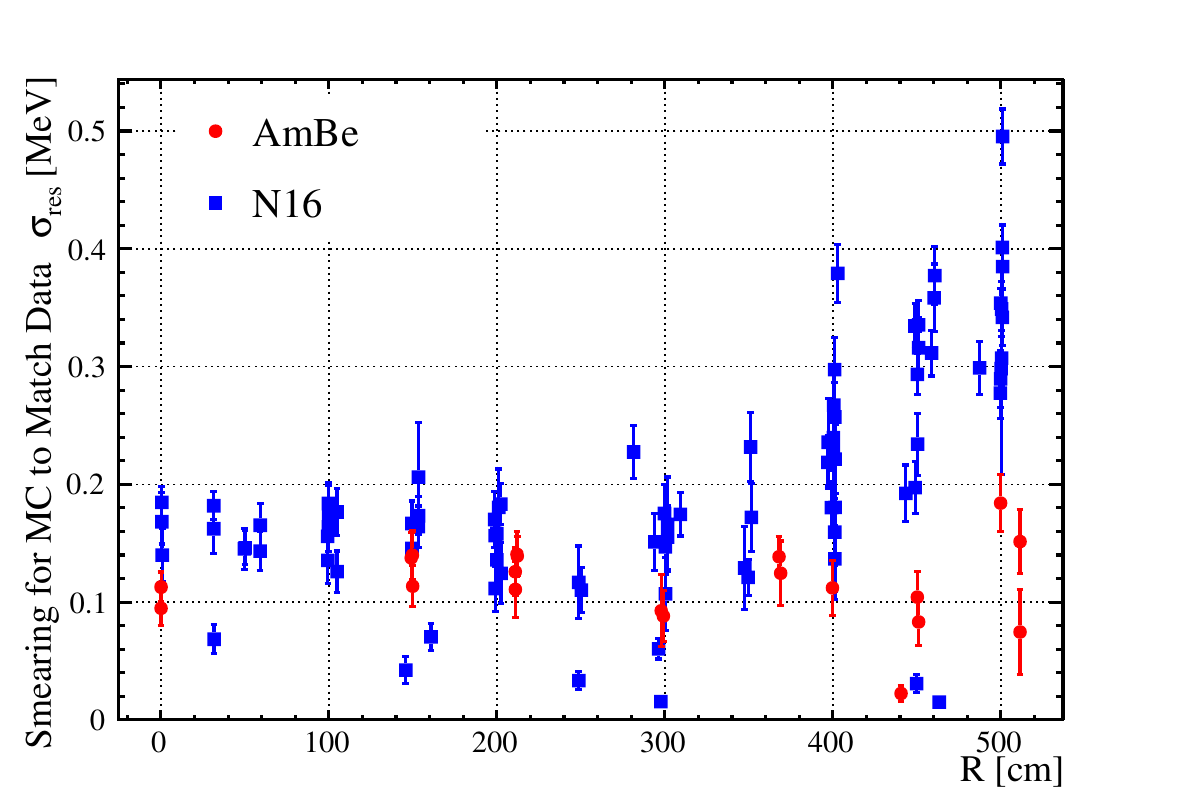}%
\caption{Comparison of the energy resolution in Data and MC as a function of calibration source radial position.}
\label{fig:Deployment_EResolution}
\end{figure}

Figure \ref{fig:Deployment_AmBe} shows a comparison of the reconstructed AmBe energy spectrum in data and Monte Carlo (MC) simulation with the source at the center of the detector.  Such comparisons were made at calibration deployment locations throughout the detector volume, as shown in Figure \ref{fig:Deployment_EScale} and Figure \ref{fig:Deployment_EResolution} for the energy scale and resolution parameters respectively, and the results combined through a volume-weighted average. The AmBe-derived systematics were smaller than those determined using the $^{16}$N source at higher energies (as described in \cite{SnoplusNDFirst}, and also shown in Figures \ref{fig:Deployment_EScale} and \ref{fig:Deployment_EResolution}). For consistency, the $^{16}$N-derived systematics (identical to those reported in \cite{SnoplusNDUpdated}) were applied at all energies in the analysis.

\section{Analysis Method}
\label{sec:Analysis}
\subsection{Signal Extraction} \label{sec:SignalExtraction}
In electron scattering interactions, the electron direction is highly correlated with the direction of the incident neutrino, so the electron scattering signal can be extracted by fitting the distribution of events in cos\xspace$\theta_{\odot}$, where $\theta_{\odot}$ is the angle between a reconstructed event direction and the vector from the Sun to the center of the detector at the time of the event. Detector backgrounds should be uncorrelated with the solar direction, and are assumed to provide a flat background in cos\xspace$\theta_{\odot}$, though this assumption is discussed in more detail in Section \ref{sec:BGShape}. For the solar flux measurement, data and simulated events passing cuts were binned in separate histograms with 40 cos\xspace$\theta_{\odot}$ bins. Below 6.0 MeV, the events were distributed in 0.5-MeV wide energy bins. Energy bins were 1-MeV wide from 6.0\,MeV to 10.0\,MeV, and a single energy bin was used between 10.0 and 15.0\,MeV.

The spectrum of solar neutrino recoil events was determined by performing a binned extended maximum likelihood fit to the cos\xspace$\theta_{\odot}$ spectrum in each individual energy bin. The magnitudes of the signal and background probability distribution functions (PDFs) were varied in the fits, with the PDF for the solar neutrino signal produced from MC, and the background assumed to be flat. Below 5\,MeV only DS-II was fitted, while above 5\,MeV the data from the two data sets was jointly fitted in each time bin.

A simultaneous fit across all energy bins, with the neutrino energy spectrum fixed to that of Winter \cite{Winters}, was carried out to determine the overall best fit solar neutrino flux across the SNO+ water phase data set.  

\subsection{Monte Carlo Simulation} \label{sec:MCGeneration}
SNO+ simulations, including the PDF used in the fit for the solar neutrino electron scattering signal, are generated using RAT, a GEANT4-based \cite{geant4} MC simulation framework that incorporates the trigger and detector conditions from specific data. The detector simulation models all relevant effects after the initial particle interaction, including Cherenkov light production, electron scattering processes, photon propagation and detection, the DAQ electronics, and the trigger system.  MC events were produced on a ``run-by-run" basis - that is, using the electronic calibration and detector settings recorded from each run to reproduce time dependent changes in the detector state in the MC, and are processed and reconstructed in the same way as data. The \beight events were generated including neutrino oscillation effects assuming the global best-fit oscillation parameters from \cite{NuMixingPars} and the neutrino production regions from \cite{GlobalSolarFlux}.

\subsection{Systematic Uncertainties}\label{sec:Systematics}
As described in Section \ref{sec:DetectorCalibration}, systematic uncertainties on reconstructed variables were assessed using comparisons of $^{16}$N calibration data to MC simulations, and confirmed in the low energy region using AmBe calibration data. These are the same as the uncertainties used in the updated optical model described in \cite{SnoplusNDUpdated}, with the addition of an absolute energy scale systematic in the spectral analysis that accounts for the uncertainty in the reconstructed energy scale relative to the true energy.

To determine the resultant uncertainties in the analysis, each systematic was independently shifted, and the analysis fits were repeated with modified PDFs, with the resulting difference from the central value each taken to be a measure of the systematic uncertainty. The uncertainties were assumed to be independent. To propagate the uncertainties on the mixing parameters in the flux result, fits were repeated using survival probability curves with the mixing parameters shifted by 1$\sigma$. Table \ref{tab:systematics} shows the contributions to the systematic uncertainty on the flux analysis, while Figure \ref{fig:JointESpectrum_3.5-15MeV} includes the total systematic uncertainty in each energy bin of the spectral analysis.

\subsubsection{Background Shape Systematic}\label{sec:BGShape}
The background in \costhetasun \xspace was assumed to be flat in the likelihood fit. However, it is possible for backgrounds with a non-uniform angular distribution in detector space to project into non-flat distributions in \costhetasun \xspace, especially in data sets which do not span a full year. To study this potential effect, the distribution of background events in detector space must be known, and the observed transient radon activity in the detector made this difficult to model, especially at lower energies. To compensate for this, a data-driven technique was used in which the observed events in the data were ``jittered" in \costhetasun \xspace by randomly re-assigning each event a different event time within the data taking period and re-calculating the \costhetasun \xspace values. This approach is expected to retain the projection of detector backgrounds in \costhetasun \xspace while roughly randomizing the directions of the solar neutrino events. The systematic effect of a possible non-flat background was then estimated by repeating the fit to the (unaltered) data multiple times using different jittered \costhetasun \xspace distributions as the background PDFs. The mean deviation of the jittered fits from the unaltered fit was taken as the related systematic uncertainty. This approach was carried out independently in the two lowest energy bins, where the signal-to-noise ratio is relatively low. Above 4.5\,MeV the low level of detector background made the technique impractical. Therefore, jittered \costhetasun \xspace distributions from $4.0<T_{e}<4.5$\, MeV were used as the background PDFs to evaluate the non-flat background systematic for all energies above 4.0\,MeV. 

\renewcommand{\arraystretch}{1.3} 
\begin{table*}
    \centering
    \begin{tabular}{c|c|c}
        \hline \hline 
        Parameter & Systematic Range & Flux Uncertainty Contribution  \\  
        \hline
        \multirow{2}{*}{x scale}  & $(x>0)~~{}^{+0.16}_{-0.23}$ \%& \\
                                       & $(x<0)~~{}^{+0.17}_{-0.30}$ \%&\\
        \multirow{2}{*}{y scale}  & $(y>0)~~{}^{+0.12}_{-0.22}$ \%&  ${}^{+0.5}_{-0.9}$\% \\
                                       & $(y<0)~~{}^{+0.17}_{-0.45}$ \%& \\
        \multirow{2}{*}{z scale}  & $(z>0)~~{}^{+0.30}_{-0.42}$ \%& \\
                                     & $(z<0)~~{}^{+0.09}_{-0.24}$ \%& \\
        \hline
        x offset & ${}^{+50.1}_{-55.6}$ mm&\\
        y offset & ${}^{+47.7}_{-59.6}$ mm& ${}^{+0.05}_{-0.01}$\%\\
        z offset & ${}^{+75.8}_{-34.7}$ mm& \\
        \hline
        x resolution & \footnotesize{$\sqrt{3214 + |- 290 + 0.393x|}$ mm}& \\
        y resolution & \footnotesize{$\sqrt{2004 + |- 1365 + 0.809y |}$ mm}& $\pm$ 0.03\%\\
        z resolution  & \footnotesize{$\sqrt{7230 + |3211 - 0.730z|}$ mm}&\\
        \hline
        Angular resolution & ${}^{+0.122}_{-0.020}$& ${}^{+1.9}_{-0.3}$\%\\
        \hline
        $\beta_{14}$ & ${}^{+0.003}_{-0.010}$& ${}^{+0.007}_{-0.02}$\%\\
        \hline
        Energy scale (Absolute) & $\pm$ 2 \% & $\pm$ 1.7\%\\
        Energy scale (Data-MC Relative) & $\pm$ 1 \% & $\pm$ 1.1\%\\
        Energy resolution & $\pm0.0084$\ & $\pm$ 0.1\%\\
        \hline
        $\theta_{21}$ ($33.02^{\circ}$ assumed) \cite{NuMixingPars}    & $^{+0.54^{\circ}}_{-0.46^{\circ}}$        & ${}^{+1.2}_{-0.9}$\%\\
        $\Delta m^{2}_{21}$ ($7.37\SI{1e-5}{\eV}^{2}$ assumed) \cite{NuMixingPars}   & $^{+0.17}_{-0.16}\times 10^{-5} \text{eV}^{2}$  & ${}^{+0.35}_{-0.33}$\%\\
        \hline
        FOM Cut Acceptance        &  $\pm
        0.2\%$    & $\pm 0.15\%$\\
        \hline
        BG Shape  & see text & ${}^{+0.7}_{-1.7}$\%\\
        \hline 
    \end{tabular} 
    \caption{
    Summary of the evaluated systematic uncertainties. The uncertainty contributions shown are for the integrated flux measurement; bin-by-bin systematics were also evaluated and are shown in Figure \ref{fig:JointESpectrum_3.5-15MeV}.
    }
    \label{tab:systematics}
\end{table*}


\section{Results} 
\label{sec:Results}
\begin{figure}[htp] 
    \centering
\includegraphics[width=0.5\textwidth]{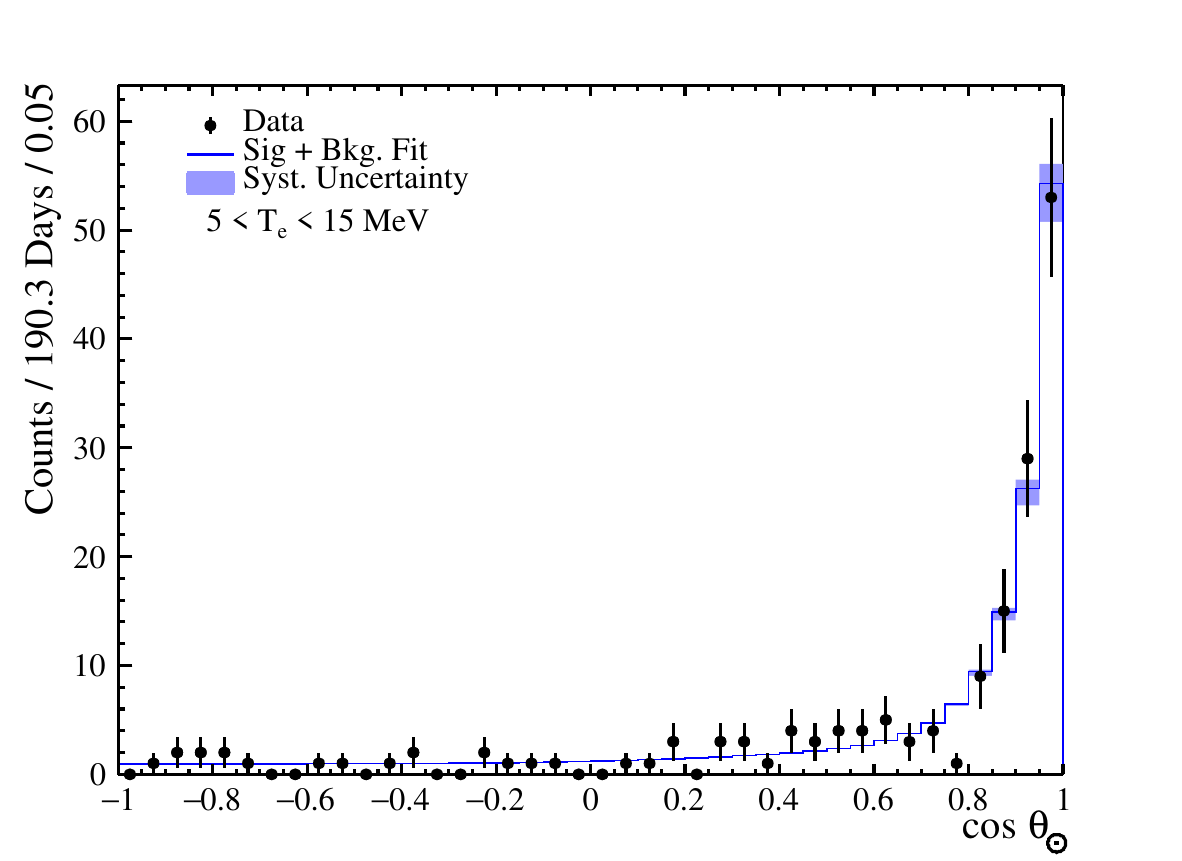}%
\caption{Distribution of event direction with respect to solar direction for DS-II
    events with energy in \numrange[range-phrase=--]{5}{15}\,MeV.}
\label{fig:5_15MeV_tb7}
\end{figure}

\begin{figure}[htbp] 
    \centering
\includegraphics[width=0.5\textwidth]{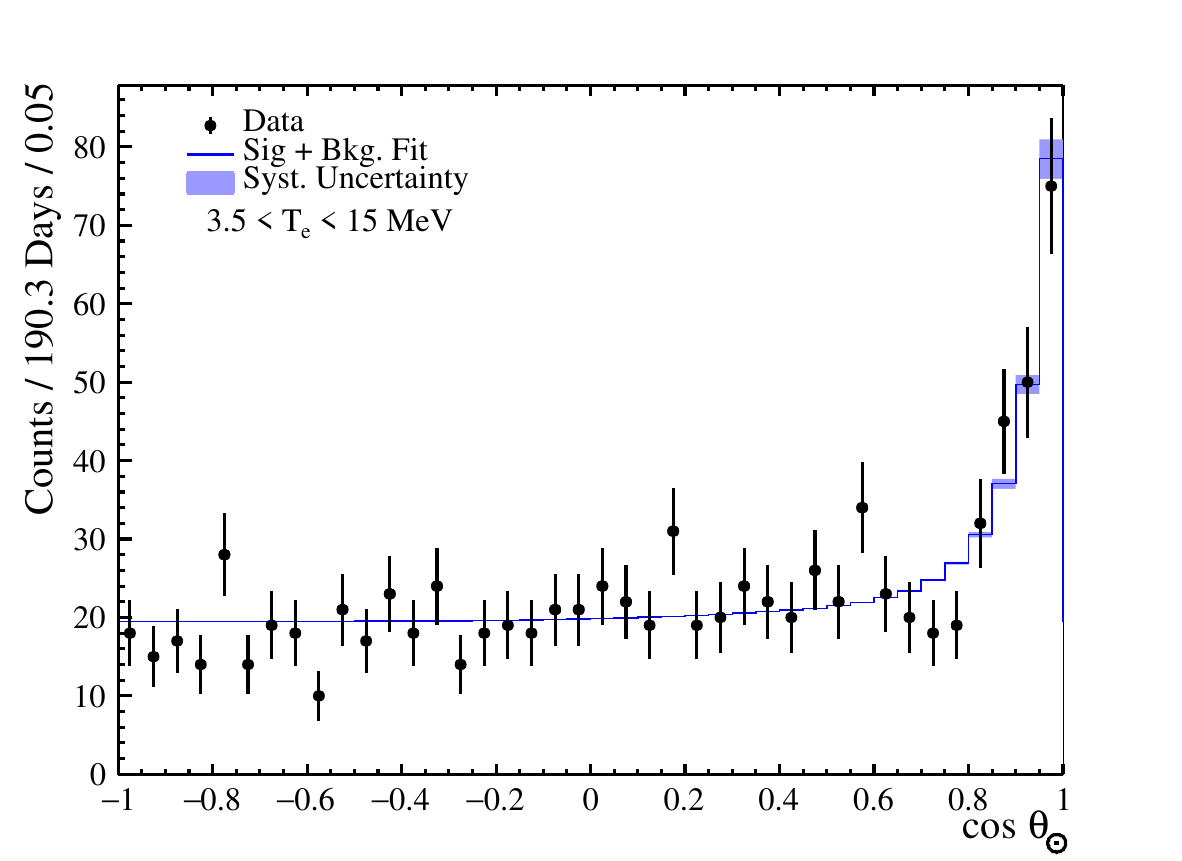}%
\caption{Distribution of event direction with respect to solar direction for DS-II events with energy in \numrange[range-phrase=--]{3.5}{15}\,MeV.}
\label{fig:3.5_15MeV_tb7}
\end{figure}

Figure \ref{fig:5_15MeV_tb7} shows the angular distribution of events above 5 MeV in the post-cover gas dataset (DS-II). The low level of background outside of the solar peak is evident; a fit to the angular distribution yields a background rate of $0.32 \pm 0.07$ events/kt-day, which is the lowest backgrounds ever measured in a water Cherenkov detector at that energy threshold. Figure \ref{fig:3.5_15MeV_tb7} shows the distribution of events in \costhetasun \xspace for events in the low background dataset (DS-II) over the entire energy range of $3.5$ to $15.0$\,MeV.

\begin{figure*}[!htbp]
    \centering
\includegraphics[width=0.5\textwidth]{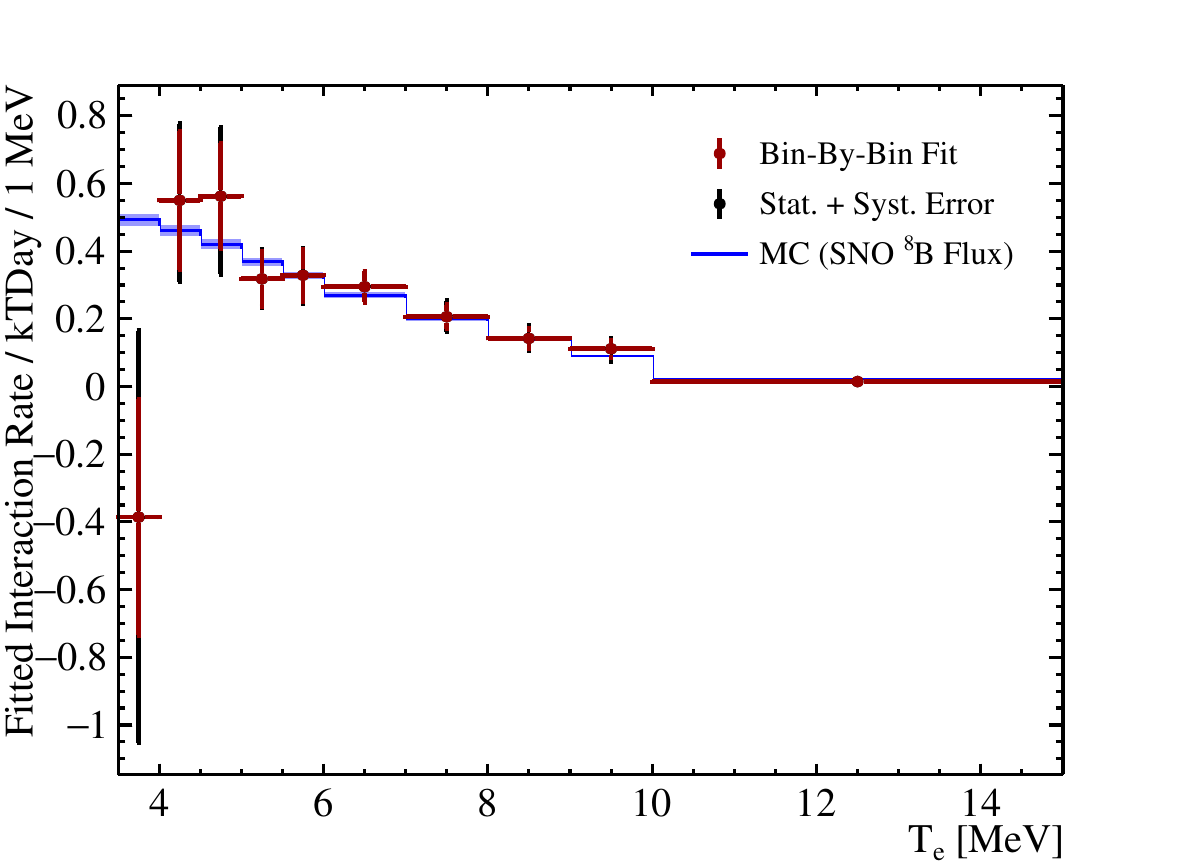}\includegraphics[width=0.5\textwidth]{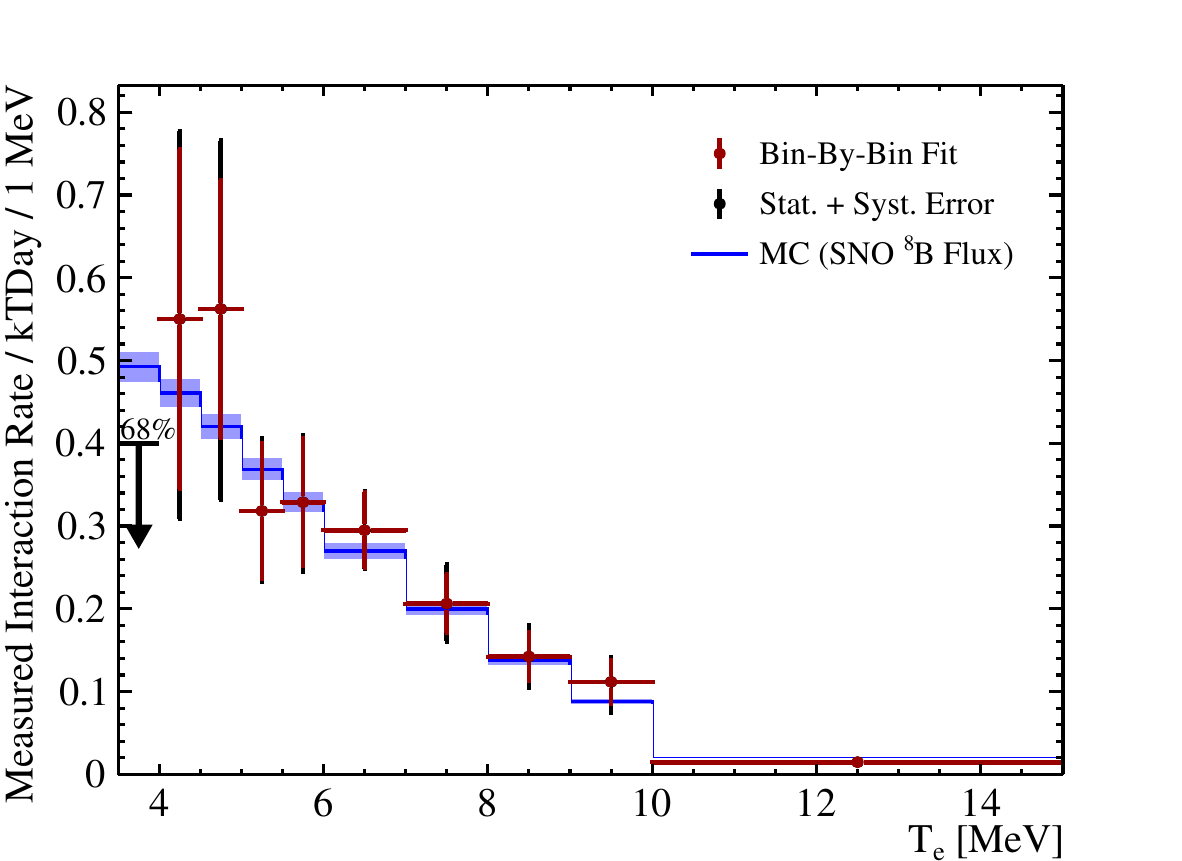}%
\caption{(Left) The fitted solar neutrino event rate and (Right) the measured neutrino interaction rate as a function of reconstructed electron kinetic energy $T_{e}$ for the joint water dataset. The lowest bin on the right represents the 68 \% C.L. Bayesian  upper limit on the interaction rate.}
\label{fig:JointESpectrum_3.5-15MeV}
\end{figure*}

Fitting the data from the two data sets simultaneously yields the solar neutrino electron recoil spectrum shown in Figure \ref{fig:JointESpectrum_3.5-15MeV}. Simultaneously fitting all energy bins in both data sets yields a combined best fit solar flux 
of \snoplusFlux. Repeating the fit under the assumption of no neutrino oscillations yields a best fit flux 
of \snoplusFluxUnoscillated. Our result is consistent with both the high metallicity and low metallicity Standard Solar Model fluxes \cite{NewSSM}.

\subsubsection{Lowest Energy Bin Discussion}\label{sec:LowestEBin}
\begin{figure}[!htb]
    \centering
\includegraphics[width=0.5\textwidth]{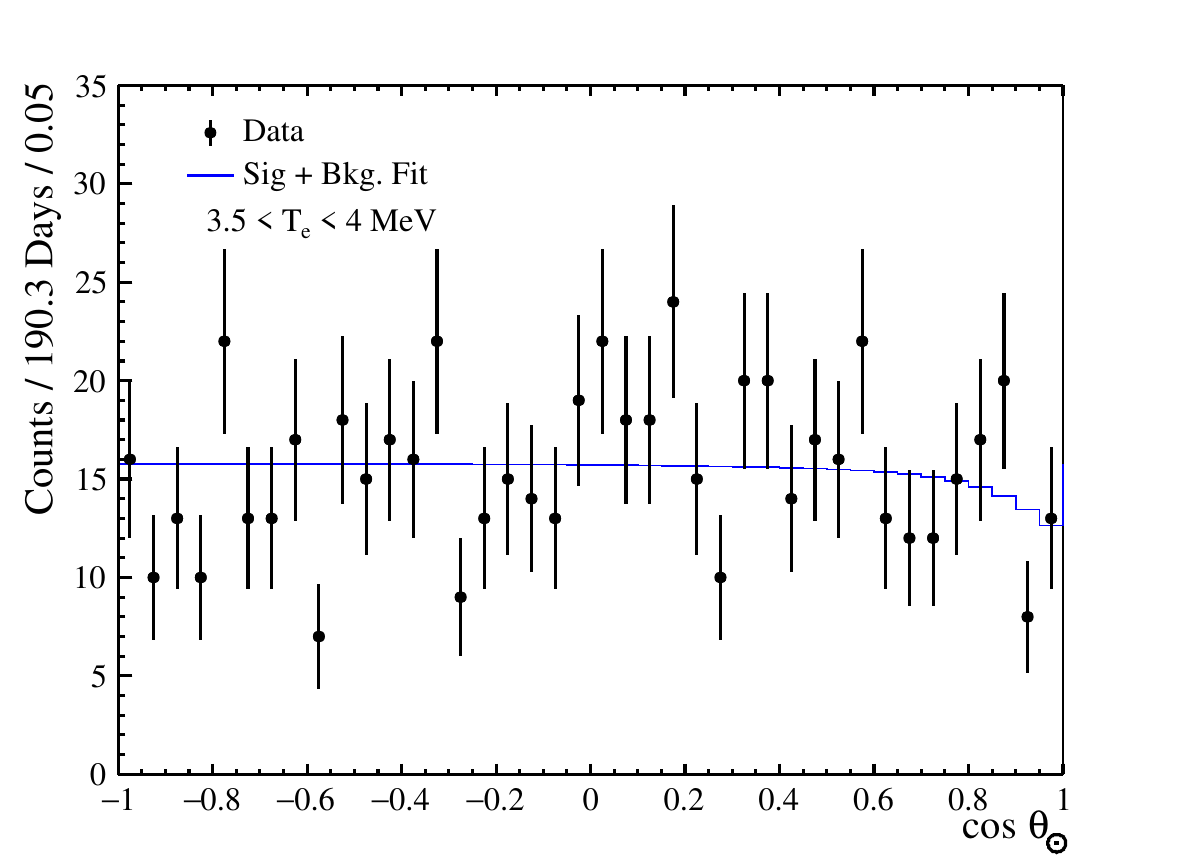}%
\caption{Distribution of event direction with respect to solar direction for DS-II
    events with energy in 3.5--4.0\, MeV.}
\label{fig:3.5_4_MeV_neg_TBII}
\end{figure}

\begin{figure}[!htbp] 
\includegraphics[width=0.5\textwidth]{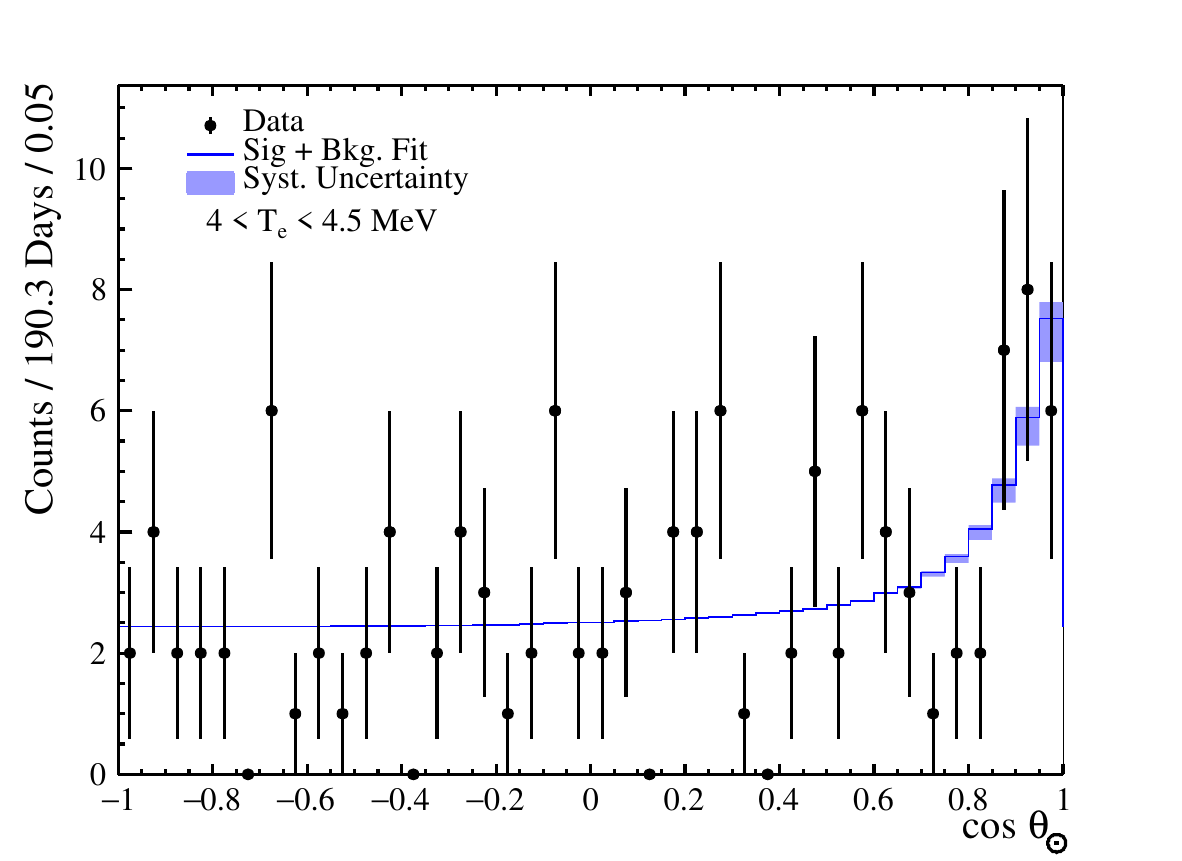}%
\caption{Fitted distributions of event direction with respect to solar direction for DS-II events with energy in 4.0--4.5\, MeV.}
\label{fig:4_4.5MeV_TBII}
\end{figure}

\begin{figure}[htbp] 
    \centering
\includegraphics[width=0.5\textwidth]{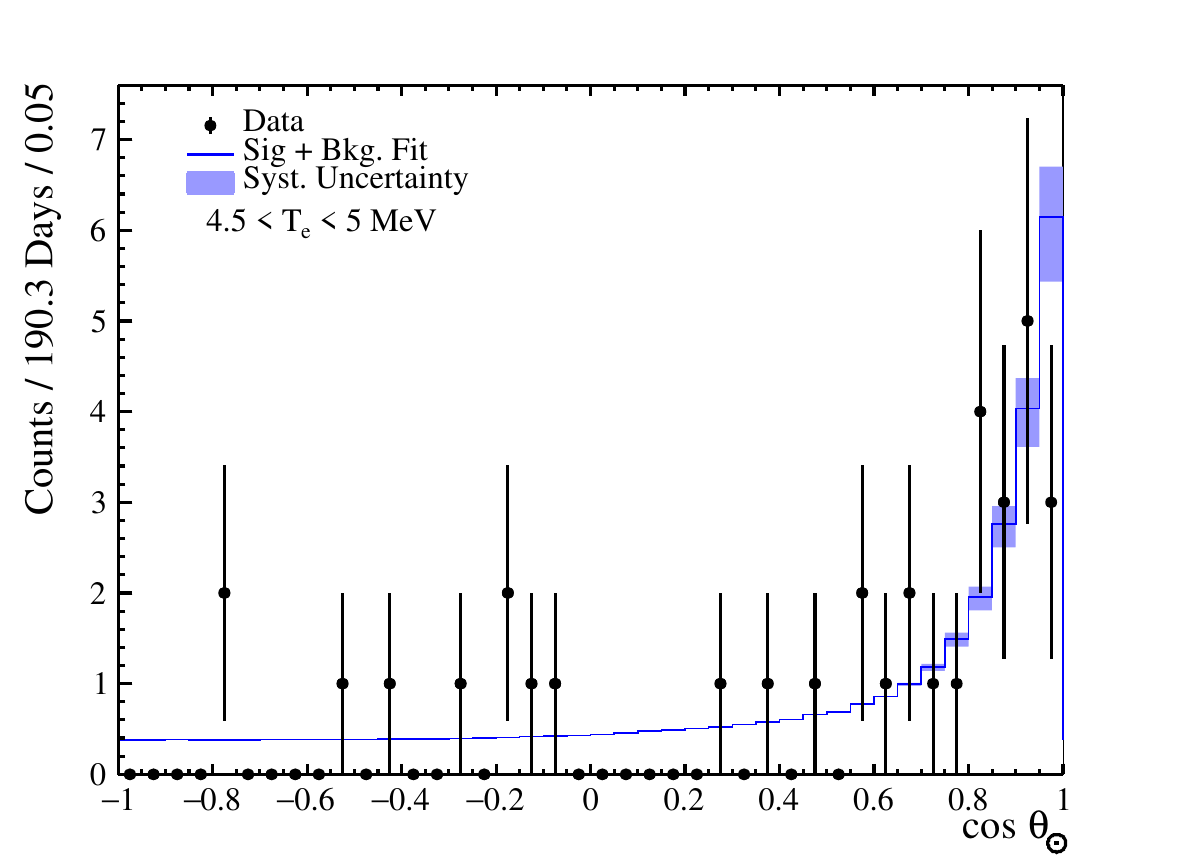}%
\caption{Distribution of event direction with respect to solar direction for DS-II events with energy in 4.5--5.0\, MeV.}
\label{fig:4.5_5MeV_TBII}
\end{figure}

As shown from Figure \ref{fig:JointESpectrum_3.5-15MeV}, the fitted ES event rate in the lowest energy (3.5 -- 4.0\,MeV) bin falls below expectation, although the uncertainties are large. The fitted \costhetasun \xspace distributions for the lowest three energy bins are shown in Figures \ref{fig:3.5_4_MeV_neg_TBII}, \ref{fig:4_4.5MeV_TBII} and \ref{fig:4.5_5MeV_TBII}.

\begin{figure}[htbp] 
    \centering
\includegraphics[width=0.5\textwidth]{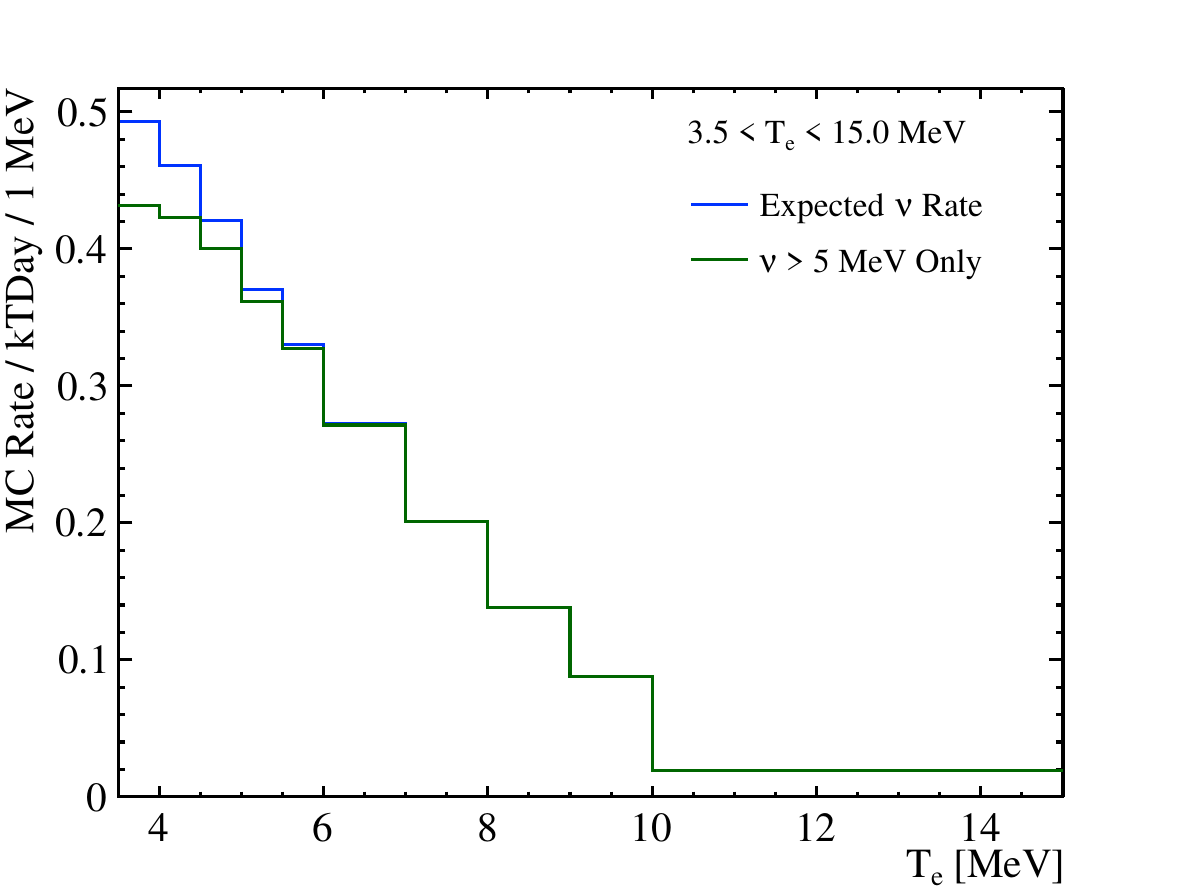}%
\caption{The MC simulated electron scatter interaction rate (blue) and with all neutrinos below 5 MeV ``turned off" (green).}
\label{fig:NuPDF_5MeVNuCut}
\end{figure}

Although the statistical significance of the deficit in the lowest energy bin is small, we feel it worthwhile to note that that the two-body kinematics of the electron scattering process means that the electron recoil signal in the low energy bins is dominated by the same higher energy neutrinos as the higher energy bins. To illustrate this effect, a MC simulation was performed with the solar neutrino flux below 5\,MeV arbitrarily set to zero. As can be seen in Figure \ref{fig:NuPDF_5MeVNuCut}, this has a rather limited effect on the expected recoil spectrum; in particular, the expected number of events in the (3.5 -- 4.0\,MeV) bin is reduced from 16.5 to 14.5. The downward fluctuation in the lowest energy bin therefore cannot be the result of a low energy distortion in the neutrino spectrum.

\section{Conclusion}
\label{sec:Conclusion}
Solar neutrino electron elastic scattering has been investigated using the full SNO+ water phase data set. The newly included data has the lowest background to the solar electron scattering signal yet demonstrated by a water Cherenkov detector, and this allowed the electron recoil spectrum to be determined down to a threshold of 3.5\,MeV. 
The measured electron recoil rate corresponds to an unoscillated solar neutrino flux of \snoplusFluxUnoscillated, or a flux of \snoplusFlux \xspace assuming standard oscillation parameters.

\begin{acknowledgments} 
Capital funds for SNO+ were provided by the Canada Foundation for Innovation and matching partners: Ontario Ministry of Research, Innovation and Science, Alberta Science and Research Investments Program, Queen's University at Kingston, and the Federal Economic Development Agency for Northern Ontario. This research was supported by (Canada) the Natural Sciences and Engineering Research Council of Canada, the Canadian Institute for Advanced Research, the Ontario Early Researcher Awards; (U.S.) the Department of Energy (DOE) Office of Nuclear Physics, the National Science Foundation, and the DOE National Nuclear Security Administration through the Nuclear Science and Security Consortium; (U.K.) the Science and Technology Facilities Council and the Royal Society; (Portugal) Fundação para a Ciência e a Tecnologia (FCT-Portugal); (Germany) the Deutsche Forschungsgemeinschaft; (Mexico) DGAPA-UNAM and Consejo Nacional de Ciencia y Tecnología; and (China) the Discipline Construction Fund of Shandong University. We also thank SNOLAB and SNO+ technical staff; the Digital Research Alliance of Canada; the GridPP Collaboration and support from Rutherford Appleton Laboratory, and the Savio computational cluster at the University of California, Berkeley. Additional long-term storage was provided by the Fermilab Scientific Computing Division. 

For the purposes of open access, the authors have applied a Creative Commons Attribution licence to any Author Accepted Manuscript version arising. Representations of the data relevant to the conclusions drawn here are provided within this paper.

\end{acknowledgments}

\bibliography{References}

\begin{thebibliography}{23}%
\makeatletter
\providecommand \@ifxundefined [1]{%
 \@ifx{#1\undefined}
}%
\providecommand \@ifnum [1]{%
 \ifnum #1\expandafter \@firstoftwo
 \else \expandafter \@secondoftwo
 \fi
}%
\providecommand \@ifx [1]{%
 \ifx #1\expandafter \@firstoftwo
 \else \expandafter \@secondoftwo
 \fi
}%
\providecommand \natexlab [1]{#1}%
\providecommand \enquote  [1]{``#1''}%
\providecommand \bibnamefont  [1]{#1}%
\providecommand \bibfnamefont [1]{#1}%
\providecommand \citenamefont [1]{#1}%
\providecommand \href@noop [0]{\@secondoftwo}%
\providecommand \href [0]{\begingroup \@sanitize@url \@href}%
\providecommand \@href[1]{\@@startlink{#1}\@@href}%
\providecommand \@@href[1]{\endgroup#1\@@endlink}%
\providecommand \@sanitize@url [0]{\catcode `\\12\catcode `\$12\catcode
  `\&12\catcode `\#12\catcode `\^12\catcode `\_12\catcode `\%12\relax}%
\providecommand \@@startlink[1]{}%
\providecommand \@@endlink[0]{}%
\providecommand \url  [0]{\begingroup\@sanitize@url \@url }%
\providecommand \@url [1]{\endgroup\@href {#1}{\urlprefix }}%
\providecommand \urlprefix  [0]{URL }%
\providecommand \Eprint [0]{\href }%
\providecommand \doibase [0]{https://doi.org/}%
\providecommand \selectlanguage [0]{\@gobble}%
\providecommand \bibinfo  [0]{\@secondoftwo}%
\providecommand \bibfield  [0]{\@secondoftwo}%
\providecommand \translation [1]{[#1]}%
\providecommand \BibitemOpen [0]{}%
\providecommand \bibitemStop [0]{}%
\providecommand \bibitemNoStop [0]{.\EOS\space}%
\providecommand \EOS [0]{\spacefactor3000\relax}%
\providecommand \BibitemShut  [1]{\csname bibitem#1\endcsname}%
\let\auto@bib@innerbib\@empty
\bibitem [{\citenamefont {Bahcall}\ \emph {et~al.}(2005)\citenamefont
  {Bahcall}, \citenamefont {Serenelli},\ and\ \citenamefont
  {Basu}}]{BS2005-OP}%
  \BibitemOpen
  \bibfield  {author} {\bibinfo {author} {\bibfnamefont {J.~N.}\ \bibnamefont
  {Bahcall}}, \bibinfo {author} {\bibfnamefont {A.~M.}\ \bibnamefont
  {Serenelli}},\ and\ \bibinfo {author} {\bibfnamefont {S.}~\bibnamefont
  {Basu}},\ }\bibfield  {title} {\bibinfo {title} {New solar opacities,
  abundances, helioseismology, and neutrino fluxes},\ }\href
  {https://doi.org/10.1086/428929} {\bibfield  {journal} {\bibinfo  {journal}
  {Astrophys. J. Lett.}\ }\textbf {\bibinfo {volume} {621}},\ \bibinfo {pages}
  {L85} (\bibinfo {year} {2005})}\BibitemShut {NoStop}%
\bibitem [{\citenamefont {Aharmim}\ \emph {et~al.}(2013)\citenamefont {Aharmim}
  \emph {et~al.}}]{SNO_LETA}%
  \BibitemOpen
  \bibfield  {author} {\bibinfo {author} {\bibfnamefont {B.}~\bibnamefont
  {Aharmim}} \emph {et~al.} (\bibinfo {collaboration} {SNO Collaboration}),\
  }\bibfield  {title} {\bibinfo {title} {{Combined analysis of all three phases
  of solar neutrino data from the Sudbury Neutrino Observatory}},\ }\href
  {https://doi.org/10.1103/PhysRevC.88.025501} {\bibfield  {journal} {\bibinfo
  {journal} {Phys. Rev. C}\ }\textbf {\bibinfo {volume} {88}},\ \bibinfo
  {pages} {025501} (\bibinfo {year} {2013})}\BibitemShut {NoStop}%
\bibitem [{\citenamefont {Abe}\ \emph {et~al.}(2024)\citenamefont {Abe} \emph
  {et~al.}}]{SuperKSolar}%
  \BibitemOpen
  \bibfield  {author} {\bibinfo {author} {\bibfnamefont {K.}~\bibnamefont
  {Abe}} \emph {et~al.} (\bibinfo {collaboration} {Super-Kamiokande
  Collaboration}),\ }\bibfield  {title} {\bibinfo {title} {{Solar neutrino
  measurements using the full data period of Super-Kamiokande-IV}},\ }\href
  {https://doi.org/10.1103/PhysRevD.109.092001} {\bibfield  {journal} {\bibinfo
   {journal} {Phys. Rev. D}\ }\textbf {\bibinfo {volume} {109}},\ \bibinfo
  {pages} {092001} (\bibinfo {year} {2024})}\BibitemShut {NoStop}%
\bibitem [{\citenamefont {Bellini}\ \emph {et~al.}(2010)\citenamefont {Bellini}
  \emph {et~al.}}]{BX:first_b8}%
  \BibitemOpen
  \bibfield  {author} {\bibinfo {author} {\bibfnamefont {G.}~\bibnamefont
  {Bellini}} \emph {et~al.} (\bibinfo {collaboration} {Borexino
  Collaboration}),\ }\bibfield  {title} {\bibinfo {title} {{Measurement of the
  solar $^{8}\mathrm{B}$ neutrino rate with a liquid scintillator target and 3
  $\mathrm{M}$e$\mathrm{V}$ energy threshold in the Borexino detector}},\
  }\href {https://doi.org/10.1103/PhysRevD.82.033006} {\bibfield  {journal}
  {\bibinfo  {journal} {Phys. Rev. D}\ }\textbf {\bibinfo {volume} {82}},\
  \bibinfo {pages} {033006} (\bibinfo {year} {2010})}\BibitemShut {NoStop}%
\bibitem [{\citenamefont {Abe}\ \emph {et~al.}(2011)\citenamefont {Abe} \emph
  {et~al.}}]{Kamland_solar}%
  \BibitemOpen
  \bibfield  {author} {\bibinfo {author} {\bibfnamefont {S.}~\bibnamefont
  {Abe}} \emph {et~al.} (\bibinfo {collaboration} {KamLAND collaboration}),\
  }\bibfield  {title} {\bibinfo {title} {{Measurement of the $^{8}\mathrm{B}$
  solar neutrino flux with the KamLAND liquid scintillator detector}},\ }\href
  {https://doi.org/10.1103/PhysRevC.84.035804} {\bibfield  {journal} {\bibinfo
  {journal} {Phys. Rev. C}\ }\textbf {\bibinfo {volume} {84}},\ \bibinfo
  {pages} {035804} (\bibinfo {year} {2011})}\BibitemShut {NoStop}%
\bibitem [{\citenamefont {Wurm}(2017)}]{Upturn_Wurm}%
  \BibitemOpen
  \bibfield  {author} {\bibinfo {author} {\bibfnamefont {M.}~\bibnamefont
  {Wurm}},\ }\bibfield  {title} {\bibinfo {title} {Solar neutrino
  spectroscopy},\ }\href {https://doi.org/10.1016/j.physrep.2017.04.002}
  {\bibfield  {journal} {\bibinfo  {journal} {Physics Reports}\ }\textbf
  {\bibinfo {volume} {685}},\ \bibinfo {pages} {1–52} (\bibinfo {year}
  {2017})}\BibitemShut {NoStop}%
\bibitem [{\citenamefont {Minakata}\ and\ \citenamefont
  {Pena-Garay}(2012)}]{Upturn_Minakata}%
  \BibitemOpen
  \bibfield  {author} {\bibinfo {author} {\bibfnamefont {H.}~\bibnamefont
  {Minakata}}\ and\ \bibinfo {author} {\bibfnamefont {C.}~\bibnamefont
  {Pena-Garay}},\ }\bibfield  {title} {\bibinfo {title} {{Solar Neutrino
  Observables Sensitive to Matter Effects}},\ }\href
  {https://doi.org/10.1155/2012/349686} {\bibfield  {journal} {\bibinfo
  {journal} {Adv. High Energy Phys.}\ }\textbf {\bibinfo {volume} {2012}},\
  \bibinfo {pages} {349686} (\bibinfo {year} {2012})}\BibitemShut {NoStop}%
\bibitem [{\citenamefont {Albanese}\ \emph {et~al.}(2021)\citenamefont
  {Albanese} \emph {et~al.}}]{SnoplusDetector}%
  \BibitemOpen
  \bibfield  {author} {\bibinfo {author} {\bibfnamefont {V.}~\bibnamefont
  {Albanese}} \emph {et~al.} (\bibinfo {collaboration} {{$\mathrm{SNO}+$
  Collaboration}}),\ }\bibfield  {title} {\bibinfo {title} {{The
  $\mathrm{SNO}+$ Experiment}},\ }\href
  {https://doi.org/10.1088/1748-0221/16/08/P08059} {\bibfield  {journal}
  {\bibinfo  {journal} {JINST}\ }\textbf {\bibinfo {volume} {16}}\bibinfo
  {number} { (08)},\ \bibinfo {pages} {P08059}}\BibitemShut {NoStop}%
\bibitem [{\citenamefont {Bahcall}(1989)}]{Bahcall_ES}%
  \BibitemOpen
\bibfield  {number} {  }\bibfield  {author} {\bibinfo {author} {\bibfnamefont
  {J.~N.}\ \bibnamefont {Bahcall}},\ }\href@noop {} {\emph {\bibinfo {title}
  {Neutrino Astrophysics}}}\ (\bibinfo  {publisher} {Cambridge University
  Press},\ \bibinfo {address} {Cambridge, UK},\ \bibinfo {year} {1989})\ pp.\
  \bibinfo {pages} {193--243}\BibitemShut {NoStop}%
\bibitem [{\citenamefont {Allega}\ \emph {et~al.}(2022)\citenamefont {Allega}
  \emph {et~al.}}]{SnoplusNDUpdated}%
  \BibitemOpen
  \bibfield  {author} {\bibinfo {author} {\bibfnamefont {A.}~\bibnamefont
  {Allega}} \emph {et~al.} (\bibinfo {collaboration} {$\mathrm{SNO}+$
  \text{Collaboration}}),\ }\bibfield  {title} {\bibinfo {title} {Improved
  search for invisible modes of nucleon decay in water with the $\mathrm{SNO}+$
  \text{detector}},\ }\href {https://doi.org/10.1103/PhysRevD.105.112012}
  {\bibfield  {journal} {\bibinfo  {journal} {Phys. Rev. D}\ }\textbf {\bibinfo
  {volume} {105}},\ \bibinfo {pages} {112012} (\bibinfo {year}
  {2022})}\BibitemShut {NoStop}%
\bibitem [{\citenamefont {Anderson}\ \emph {et~al.}(2020)\citenamefont
  {Anderson} \emph {et~al.}}]{SnoplusNeutronProton}%
  \BibitemOpen
  \bibfield  {author} {\bibinfo {author} {\bibfnamefont {M.~R.}\ \bibnamefont
  {Anderson}} \emph {et~al.} (\bibinfo {collaboration} {$\mathrm{SNO}+$
  Collaboration}),\ }\bibfield  {title} {\bibinfo {title} {Measurement of
  neutron-proton capture in the $\mathrm{SNO}+$ water phase},\ }\href
  {https://doi.org/10.1103/PhysRevC.102.014002} {\bibfield  {journal} {\bibinfo
   {journal} {Phys. Rev. C}\ }\textbf {\bibinfo {volume} {102}},\ \bibinfo
  {pages} {014002} (\bibinfo {year} {2020})}\BibitemShut {NoStop}%
\bibitem [{\citenamefont {Allega}\ \emph {et~al.}(2023)\citenamefont {Allega}
  \emph {et~al.}}]{SnoplusAntinu}%
  \BibitemOpen
  \bibfield  {author} {\bibinfo {author} {\bibfnamefont {A.}~\bibnamefont
  {Allega}} \emph {et~al.} (\bibinfo {collaboration} {$\mathrm{SNO}+$
  Collaboration}),\ }\bibfield  {title} {\bibinfo {title} {Evidence of
  antineutrinos from distant reactors using pure water at $\mathrm{SNO}+$},\
  }\href {https://doi.org/10.1103/PhysRevLett.130.091801} {\bibfield  {journal}
  {\bibinfo  {journal} {Phys. Rev. Lett.}\ }\textbf {\bibinfo {volume} {130}},\
  \bibinfo {pages} {091801} (\bibinfo {year} {2023})}\BibitemShut {NoStop}%
\bibitem [{\citenamefont {Anderson}\ \emph
  {et~al.}(2019{\natexlab{a}})\citenamefont {Anderson} \emph
  {et~al.}}]{SnoplusPrior8B}%
  \BibitemOpen
  \bibfield  {author} {\bibinfo {author} {\bibfnamefont {M.}~\bibnamefont
  {Anderson}} \emph {et~al.} (\bibinfo {collaboration} {$\mathrm{SNO}+$
  Collaboration}),\ }\bibfield  {title} {\bibinfo {title} {Measurement of the
  $^{8}\mathrm{B}$ solar neutrino flux in $\mathrm{SNO}+$ with very low
  backgrounds},\ }\href {https://doi.org/10.1103/PhysRevD.99.012012} {\bibfield
   {journal} {\bibinfo  {journal} {Phys. Rev. D}\ }\textbf {\bibinfo {volume}
  {99}},\ \bibinfo {pages} {012012} (\bibinfo {year}
  {2019}{\natexlab{a}})}\BibitemShut {NoStop}%
\bibitem [{\citenamefont {Boger}\ \emph {et~al.}(2000)\citenamefont {Boger}
  \emph {et~al.}}]{SNO_NIM}%
  \BibitemOpen
  \bibfield  {author} {\bibinfo {author} {\bibfnamefont {J.}~\bibnamefont
  {Boger}} \emph {et~al.} (\bibinfo {collaboration} {SNO Collaboration}),\
  }\bibfield  {title} {\bibinfo {title} {{The Sudbury Neutrino Observatory}},\
  }\href {https://doi.org/https://doi.org/10.1016/S0168-9002(99)01469-2}
  {\bibfield  {journal} {\bibinfo  {journal} {Nucl. Instr. And Meth. A}\
  }\textbf {\bibinfo {volume} {449}},\ \bibinfo {pages} {172} (\bibinfo {year}
  {2000})}\BibitemShut {NoStop}%
\bibitem [{\citenamefont {Anderson}\ \emph
  {et~al.}(2019{\natexlab{b}})\citenamefont {Anderson} \emph
  {et~al.}}]{SnoplusNDFirst}%
  \BibitemOpen
  \bibfield  {author} {\bibinfo {author} {\bibfnamefont {M.}~\bibnamefont
  {Anderson}} \emph {et~al.} (\bibinfo {collaboration} {{$\mathrm{SNO}+$
  Collaboration}}),\ }\bibfield  {title} {\bibinfo {title} {{Search for
  invisible modes of nucleon decay in water with the $\mathrm{SNO}+$
  detector}},\ }\href {https://doi.org/10.1103/PhysRevD.99.032008} {\bibfield
  {journal} {\bibinfo  {journal} {Phys. Rev. D}\ }\textbf {\bibinfo {volume}
  {99}},\ \bibinfo {pages} {032008} (\bibinfo {year}
  {2019}{\natexlab{b}})}\BibitemShut {NoStop}%
\bibitem [{\citenamefont {Ahmed}\ \emph {et~al.}(2004)\citenamefont {Ahmed}
  \emph {et~al.}}]{SNO_Salt}%
  \BibitemOpen
  \bibfield  {author} {\bibinfo {author} {\bibfnamefont {S.~N.}\ \bibnamefont
  {Ahmed}} \emph {et~al.} (\bibinfo {collaboration} {SNO Collaboration}),\
  }\bibfield  {title} {\bibinfo {title} {{Measurement of the Total Active
  $^{8}\mathrm{B}$ Solar Neutrino Flux at the Sudbury Neutrino Observatory with
  Enhanced Neutral Current Sensitivity}},\ }\href
  {https://doi.org/10.1103/PhysRevLett.92.181301} {\bibfield  {journal}
  {\bibinfo  {journal} {Phys. Rev. Lett.}\ }\textbf {\bibinfo {volume} {92}},\
  \bibinfo {pages} {181301} (\bibinfo {year} {2004})}\BibitemShut {NoStop}%
\bibitem [{\citenamefont {Anderson}\ \emph {et~al.}(2021)\citenamefont
  {Anderson} \emph {et~al.}}]{SnoplusOptics}%
  \BibitemOpen
  \bibfield  {author} {\bibinfo {author} {\bibfnamefont {M.}~\bibnamefont
  {Anderson}} \emph {et~al.} (\bibinfo {collaboration} {{$\mathrm{SNO}+$
  Collaboration}}),\ }\bibfield  {title} {\bibinfo {title} {{Optical
  calibration of the $\mathrm{SNO}+$ detector in the water phase with deployed
  sources}},\ }\href
  {https://doi.org/https://doi.org/10.1088/1748-0221/16/10/P10021} {\bibfield
  {journal} {\bibinfo  {journal} {JINST}\ }\textbf {\bibinfo {volume}
  {16}}\bibinfo  {number} { (10)},\ \bibinfo {pages} {P10021}}\BibitemShut
  {NoStop}%
\bibitem [{\citenamefont {Aharmim}\ \emph {et~al.}(2005)\citenamefont {Aharmim}
  \emph {et~al.}}]{SNO_8B}%
  \BibitemOpen
\bibfield  {number} {  }\bibfield  {author} {\bibinfo {author} {\bibfnamefont
  {B.}~\bibnamefont {Aharmim}} \emph {et~al.} (\bibinfo {collaboration} {SNO
  Collaboration}),\ }\bibfield  {title} {\bibinfo {title} {{Electron energy
  spectra, fluxes, and day-night asymmetries of $^{8}\mathrm{B}$ solar
  neutrinos from measurements with NaCl dissolved in the heavy-water detector
  at the Sudbury Neutrino Observatory}},\ }\href
  {https://doi.org/10.1103/PhysRevC.72.055502} {\bibfield  {journal} {\bibinfo
  {journal} {Phys. Rev. C}\ }\textbf {\bibinfo {volume} {72}},\ \bibinfo
  {pages} {055502} (\bibinfo {year} {2005})}\BibitemShut {NoStop}%
\bibitem [{\citenamefont {Winter}\ \emph {et~al.}(2006)\citenamefont {Winter},
  \citenamefont {Freedman}, \citenamefont {Rehm},\ and\ \citenamefont
  {Schiffer}}]{Winters}%
  \BibitemOpen
  \bibfield  {author} {\bibinfo {author} {\bibfnamefont {W.~T.}\ \bibnamefont
  {Winter}}, \bibinfo {author} {\bibfnamefont {S.~J.}\ \bibnamefont
  {Freedman}}, \bibinfo {author} {\bibfnamefont {K.~E.}\ \bibnamefont {Rehm}},\
  and\ \bibinfo {author} {\bibfnamefont {J.~P.}\ \bibnamefont {Schiffer}},\
  }\bibfield  {title} {\bibinfo {title} {The $^{8}\mathrm{B}$ neutrino
  spectrum},\ }\href {https://doi.org/10.1103/PhysRevC.73.025503} {\bibfield
  {journal} {\bibinfo  {journal} {Phys. Rev. C}\ }\textbf {\bibinfo {volume}
  {73}},\ \bibinfo {pages} {025503} (\bibinfo {year} {2006})}\BibitemShut
  {NoStop}%
\bibitem [{\citenamefont {Agostinelli}\ \emph {et~al.}(2003)\citenamefont
  {Agostinelli} \emph {et~al.}}]{geant4}%
  \BibitemOpen
  \bibfield  {author} {\bibinfo {author} {\bibfnamefont {S.}~\bibnamefont
  {Agostinelli}} \emph {et~al.} (\bibinfo {collaboration} {{$\mathrm{GEANT4}$
  Collaboration}}),\ }\bibfield  {title} {\bibinfo {title} {Geant4---a
  simulation toolkit},\ }\href
  {https://doi.org/https://doi.org/10.1016/S0168-9002(03)01368-8} {\bibfield
  {journal} {\bibinfo  {journal} {Nucl. Instr. And Meth. A}\ }\textbf {\bibinfo
  {volume} {506}},\ \bibinfo {pages} {250 } (\bibinfo {year}
  {2003})}\BibitemShut {NoStop}%
\bibitem [{\citenamefont {Capozzi}\ \emph {et~al.}(2016)\citenamefont
  {Capozzi}, \citenamefont {Lisi}, \citenamefont {Marrone}, \citenamefont
  {Montanino},\ and\ \citenamefont {Palazzo}}]{NuMixingPars}%
  \BibitemOpen
  \bibfield  {author} {\bibinfo {author} {\bibfnamefont {F.}~\bibnamefont
  {Capozzi}}, \bibinfo {author} {\bibfnamefont {E.}~\bibnamefont {Lisi}},
  \bibinfo {author} {\bibfnamefont {A.}~\bibnamefont {Marrone}}, \bibinfo
  {author} {\bibfnamefont {D.}~\bibnamefont {Montanino}},\ and\ \bibinfo
  {author} {\bibfnamefont {A.}~\bibnamefont {Palazzo}},\ }\bibfield  {title}
  {\bibinfo {title} {Neutrino masses and mixings: Status of known and unknown 3
  $\nu$ parameters},\ }\href
  {https://doi.org/https://doi.org/10.1016/j.nuclphysb.2016.02.016} {\bibfield
  {journal} {\bibinfo  {journal} {Nuclear Physics B}\ }\textbf {\bibinfo
  {volume} {908}},\ \bibinfo {pages} {218} (\bibinfo {year}
  {2016})}\BibitemShut {NoStop}%
\bibitem [{\citenamefont {Bergstr{\"o}m}\ \emph {et~al.}(2016)\citenamefont
  {Bergstr{\"o}m}, \citenamefont {Gonzalez-Garcia}, \citenamefont {Maltoni},
  \citenamefont {Pe{\~{n}}a-Garay}, \citenamefont {Serenelli},\ and\
  \citenamefont {Song}}]{GlobalSolarFlux}%
  \BibitemOpen
  \bibfield  {author} {\bibinfo {author} {\bibfnamefont {J.}~\bibnamefont
  {Bergstr{\"o}m}}, \bibinfo {author} {\bibfnamefont {M.~C.}\ \bibnamefont
  {Gonzalez-Garcia}}, \bibinfo {author} {\bibfnamefont {M.}~\bibnamefont
  {Maltoni}}, \bibinfo {author} {\bibfnamefont {C.}~\bibnamefont
  {Pe{\~{n}}a-Garay}}, \bibinfo {author} {\bibfnamefont {A.~M.}\ \bibnamefont
  {Serenelli}},\ and\ \bibinfo {author} {\bibfnamefont {N.}~\bibnamefont
  {Song}},\ }\bibfield  {title} {\bibinfo {title} {Updated determination of the
  solar neutrino fluxes from solar neutrino data},\ }\href
  {https://doi.org/10.1007/JHEP03(2016)132} {\bibfield  {journal} {\bibinfo
  {journal} {J. High Energ. Phys}\ }\textbf {\bibinfo {volume} {2016}},\
  \bibinfo {pages} {132} (\bibinfo {year} {2016})}\BibitemShut {NoStop}%
\bibitem [{\citenamefont {Gonzalez-Garcia}\ \emph {et~al.}(2024)\citenamefont
  {Gonzalez-Garcia}, \citenamefont {Maltoni}, \citenamefont {Pinheiro},\ and\
  \citenamefont {Serenelli}}]{NewSSM}%
  \BibitemOpen
  \bibfield  {author} {\bibinfo {author} {\bibfnamefont {M.~C.}\ \bibnamefont
  {Gonzalez-Garcia}}, \bibinfo {author} {\bibfnamefont {M.}~\bibnamefont
  {Maltoni}}, \bibinfo {author} {\bibfnamefont {J.~P.}\ \bibnamefont
  {Pinheiro}},\ and\ \bibinfo {author} {\bibfnamefont {A.~M.}\ \bibnamefont
  {Serenelli}},\ }\bibfield  {title} {\bibinfo {title} {{Status of direct
  determination of solar neutrino fluxes after Borexino}},\ }\href
  {http://dx.doi.org/10.1007/JHEP02(2024)064} {\bibfield  {journal} {\bibinfo
  {journal} {J. High Energ. Phys}\ }\textbf {\bibinfo {volume} {2024}},\
  \bibinfo {pages} {64} (\bibinfo {year} {2024})}\BibitemShut {NoStop}%
\end{thebibliography}%

\end{document}